\documentclass[a4,12pt]{article}
\begin{document}
\newcommand{\al}{\alpha}
\newcommand{\be}{\beta}
\newcommand{\si}{\sigma}
\newcommand{\de}{\delta}
\newcommand{\ga}{\gamma}
\newcommand{\ep}{\epsilon}
\newcommand{\la}{\lambda}
\newcommand{\La}{\Lambda}
\newcommand{\dL}{\partial\La}
\newcommand{\ind}{\mathcal{L}}
\newcommand{\inter}{\mathcal{I}}
\newcommand{\no}{\mathcal{N}}
\newcommand{\states}{\mathcal{S}}
\newcommand{\un}{\underline{n}}
\newcommand{\ra}{\rightarrow}
\newcommand{\Ra}{\Rightarrow}
\newcommand{\da}{\downarrow}
\newcommand{\om}{\omega}
\newcommand{\ze}{\zeta}
\newcommand{\xsi}{\xi}
\newcommand{\tr}{\mbox{tr}}
\newcommand{\Tr}{\mbox{Tr}}
\newcommand{\bfe}{{\bf e}}
\newcommand{\bfE}{{\bf E}}
\newcommand{\bfP}{{\bf P}}
\newcommand{\N}{{\bf N}}
\newcommand{\R}{{\bf R}}
\newcommand{\limNinf}{\lim_{N\ra\infty}}
\newcommand{\zv}{{\vec z}}
\newcommand{\tv}{{\vec t}}
\newcommand{\xv}{{\vec x}}
\newcommand{\qv}{{\vec q}}
\newcommand{\alv}{{\vec \al}}
\newcommand{\bfeplane}{$(\bfe_1,~\bfe_2,~\bfe_3)$-plane}
\newcommand{\ls}{\la^\star}
\newcommand{\rs}{r^\star}
\newcommand{\rh}{\hat{r}}
\newcommand{\rt}{{\tilde r}}
\newcommand{\gt}{{\tilde g}}
\newcommand{\fNb}{{\bar f}_N^\star}
\newcommand{\fNbt}{{\bar f}_N}
\newcommand{\EfN}{\bfE f_N}
\newcommand{\gnt}{g_N(\theta)}
\newcommand{\gntmin}{\mbox{min}_\theta\gnt}
\newcommand{\wh}{\setminus}
\newcommand{\nin}{\in\!\!\!/} 
\newtheorem{axioma}{Axioma}[section]
\newtheorem{lemma}{Lemma}[section]
\newtheorem{theorem}{Theorem}[section]

\title{Infinitely many states and stochastic 
\break
symmetry in a Gaussian Potts-Hopfield model}
\author{A.C.D. van Enter and H.G. Schaap\footnote{Instute for Theoretical Physics, Rijksuniversiteit Groningen, NL-9747 AG Groningen, The Netherlands; e-mail: A.C.D.van.Enter@phys.rug.nl, H.G.Schaap@phys.rug.nl}}
\maketitle
\section{Introduction}
\noindent
In this paper we study the mean-field Potts model with Hopfield-Mattis 
disorder, in particular with Gaussianly distributed disorder. This model is a 
generalization of the model studied in \cite{sosy}. It provides 
yet another example of a disordered model with infinitely many
low-temperature pure states, such as is sometimes believed to be typical 
for spin-glasses \cite{MPV}. In our model, however, in contrast to 
\cite{sosy}, instead of chaotic 
pairs we find that the chaotic size dependence is realized by chaotic 
$q(q-1)$-tuples. For the notion of chaotic size dependence, and the notion of chaotic pairs
which were introduced by Newman and Stein we refer to \cite{N&S, NS2} and references mentioned there. Compare also \cite{sosy} and \cite{nie}.
For an extensive discussion of the Hopfield model, including some 
history and its relation with the theory of neural networks, see   
\cite[pag. 133 and further]{lecturenotes} or \cite{Hopfield}.
A somewhat different generalization of the Hopfield model to 
Potts spins can be found in \cite{G}.\\\\
We are concerned 
in particular with the infinite-volume limit behaviour of the Gibbs and 
ground state measures. The possible limit points are labeled as 
the minima of an appropriate mean-field (free) energy functional. 
These minima can be obtained as  
solutions of a suitable mean-field equation.
These minima lie on the minimal-free-energy surface, which is a 
$m(q-1)$-sphere in the 
$(\bfe^1,\cdots,\bfe^q)^{\otimes m}$ space. This space for $q$-state 
Potts spins and $m$ patterns is formed by the m-fold product of the 
hyperplane spanned by the end 
points of the unit vectors $\bfe_q$, which are the possible 
%outcomes 
values
of the 
spins. But only a limited area of the minimal free energy surface is 
accessible. Only those values 
for which certain mean-field equations hold, are allowed. These equations have the structure of fixed point equations. We 
derive them in chapter \ref{radius}. To obtain the 
%ground states, combine the 
Gibbs states we need to find the solutions of these equations
on the minimal free energy surface. \\\\
The structure of the ground or Gibbs states for $q=2$ and $\xi^k$ Gaussian with $m=2$ 
is known since a few years \cite{sosy}. Due to the Gaussian distribution we 
have a nice symmetric structure: the ground states form a circle. 
For a fixed configuration and a large finite volume
the possible order-parameter values become close to
two diametrical points (which ones depend on the volume of the system) 
on this circle. 
This paper 
%is about 
treats the 
generalization of this structure 
%for 
to
$q$-state Potts spins with $q>2$. To have a concrete example, 
we concentrate on the case $q=3$. It turns out that we again obtain a 
circle symmetry but also a  discrete symmetry, which 
%different for 
generalizes
the one for Ising spins. One gets 
instead of a single 
pair a triple of pairs (living on 3 separate circles), 
where for each pair one has a similar structure as for the 
single pair for $q=2$. For $q>3$ we get $q(q-1)\over 2$ pairs and a similar 
higher-dimensional structure.\\\\
Our model displays quenched disorder.
This means that we look at a 
fixed, particular realization
of the patterns. It turns out 
%to be 
that
there is some kind of 
self-averaging. The thermodynamic behaviour of the Hamiltonian is the same 
for almost every 
realization. This is the case for the free energy and the associated 
fixed point 
equations, as is familiar from many quenched disordered models. 
%But 
However,  this is not precisely
true for the order parameters. We will see that they show a form of chaotic 
size dependence, i.e. the behaviour strongly depends both on the chosen 
configuration and on the way one takes the infinite-volume limit 
$N\ra\infty$ (that is, along which subsequence).

\section{Notations and definitions}

We start with some definitions. Consider the set 
$\La_N=\{1,\cdots,N\}\in{\bf N}^+$. Let the single-spin space $\chi$ be a 
finite set and the N-spin configuration space be $\chi^{\otimes N}$. 
We denote a spin configuration 
by $\si$ and its value at site i by $\si_i$. We will consider Potts spins, in 
the Wu representation \cite{Potts}. The set $\chi^{\otimes N}$ is then the 
$N$-fold tensor product of the set $\chi=\{\bfe^1,\cdots,\bfe^q\}$. 
The $\bfe^\si$ are the projection of the spinvectors $\bfe_\si$ on the hypertetrahedron in ${\bf R}^{q-1}$ spanned by the end points of $\bfe_\si$. For $q=3$ we get for 
example for $\bfe^1$, $\bfe^2$ and $\bfe^3$ the vectors:
\[\left\{\left(\begin{array}{c}
                 1 \\
                 0 
              \end{array}\right),
                \left(\begin{array}{c}
                 -\frac{1}{2} \\
                \frac{1}{2}\sqrt{3} 
              \end{array}\right),
                \left(\begin{array}{c}
                 -\frac{1}{2} \\
                -\frac{1}{2}\sqrt{3}
              \end{array}\right)\right\}.\]
The Hamiltonian of our model is defined as follows:
\[-\be H_N=\frac{\be}{N}\sum_{k=1}^m\sum_{i,j=1}^N\xi_i^k\xi_j^k\de(\si_i,\si_j)\mbox{, with}\]
\[\de(\si_i,\si_j)=\frac{1}{q}\left[1+(q-1)\bfe^{\si_i}\cdot\bfe^{\si_j}\right],\]
where $\xi_i^k$ is the $i$-th component of the random $N$-component 
vector $\xi^k$. For the $\xi_i^k$'s we choose i.i.d. $N(0,1)$ distributions. 
The vectors $\xi^k=(\xi_1^k,\cdots,\xi_N^k)$, by analogy with the standard Hopfield model, are called patterns. If we combine the above, we can rewrite the Hamiltonian $H_N$ as: 
\[-\be H_N=\be\frac{q-1}{q}N\sum_{k=1}^m\left[\left(\frac{\sum\xi_i^k\bfe^{\si_i}}{N}\right)^2+\frac{1}{q-1}\left(\frac{\sum\xi_i^k}{N}\right)^2\right].\]
So asymptotically
\begin{equation}\label{HHopfield}
-\be H_N=N\frac{K}{2}\sum_{k=1}^mq_{kN}^2, 
\end{equation}
\[\mbox{with }K=2\be\left(\frac{q-1}{q}\right)\mbox{ and order 
parameters }q_{kN}=\frac{1}{N}\sum_{i=1}^N\xi_i^k\bfe^{\si_i}.\]
The last term is an irrelevant constant; in fact it approaches zero, 
due to the strong law of large numbers. 
(The $\xi_i^k$'s are i.i.d. $N(0,1)$ distributed so $E\xi_i^k=0$.) 
Note that any i.i.d. distribution with zero mean, finite variance and 
symmetrically distributed around zero will give an analogous form of $H_N$, 
but we plan to consider only Gaussian distributions, for which we find that 
a continous symmetry can be stochastically broken, just as 
in \cite{sosy}.  From now on we drop the 
subscript $N$ to simplify the notation, when no confusion can arise.\\\\
Furthermore we introduce two representations 
%of 
for
the order parameters 
$\qv$. If we assume $m=2$ then $\qv=(\qv_1,\qv_2)$ and the definitions are 
as follows: if we consider the space $\bf{R}^{q-1}$ 
%on which the 
spanned by the
vectors $\bfe^1,\cdots,\bfe^q$ 
%are defined, 
, the $\vec{x}$-plane, we define 
$\qv=(x_1,\cdots,x_{2(q-1)})$. It is often more convenient to look at the 
(higher-dimensional) $(\bfe_1,\cdots,\bfe_q)$-space.
%-plane. 
%Then 
In that case
we take $\qv=(a_1,a_2,a_3,b_1,b_2,b_3)$ for $q=3$ and an equivalent 
equation for other values of $q$. For $m\neq 2$ the definitions are analogous.

\section{Ground states} 
Now it is time to reveal the characteristics of the ground states for the 
Potts model. First we discuss the simple behaviour for 1 pattern. 
Then the more interesting part: $q>2$ and 2 patterns.
\subsection{Ground states for 1 pattern}
For one pattern $\xi$ the Hamiltonian is 
%in 
of
the following form:
\[-\be H=N\frac{K}{2}\qv_1^2=\frac{\be}{N}\sum_{i,j=1}^N\xi_i\xi_j\de(\si_i,\si_j).\]
We 
%can 
easily see that the ground states are obtained by directing the spins 
with $\xi_i>0$ in one direction and the spins with $\xi_i\leq 0$ 
in a different direction.
If we have as the distribution for the $\xi_i$'s $P(\xi_i=\pm 1)=\frac{1}{2}$
{\bf,} 
then the order parameter is of the form: 
$\qv_1=\frac{1}{2}(\bfe^{\si_i}-\bfe^{\si_j})$, 
with $1\leq i,~j\leq q$ and $i\neq j$, see also \cite{colour}. So for $q=3$ we have only 6 ground states. They form a regular hexagon: 
$(\pm\frac{3}{4},\mp\frac{\sqrt{3}}{4}),~\pm(\frac{3}{4},\frac{\sqrt{3}}{4}),~(0,\pm\frac{\sqrt{3}}{2})$. 
This regular hexagon with its interior is the convex set of possible 
order parameter values. It is easy to see that for $\xi_i$ 
$N(0,1)$-distributed we get the same ground states except for a scaling 
factor $\sqrt{2/\pi}$ multiplying the values of the order parameter values.

\subsection{Ground states for 2 patterns}

The Hamiltonian for 2 patterns (Gaussian i.i.d.) is: 
\[-\be H_N=\frac{\be}{N}\sum_{i,j=1}^N(\xi_i\xi_j+\eta_i\eta_j)\de(\si_i,\si_j)=N\frac{K}{2}(\qv_1^2+\qv_2^2).\]
Similarly as in \cite{sosy}, we make use of the fact that
the distribution of 2 independent 
identically distributed Gaussians has a continous rotation symmetry. 
This symmetry shows also up in the order parameters.
Let 
\begin{equation}\label{multiplication}
\qv_1(\theta)=
\left(\begin{array}{c}
          x_1(\theta) \\
          x_2(\theta) \\
      \end{array}\right)=\left(
      \begin{array}{c}
          \al\sin{\theta} \\
          \be\sin{\theta} \\
      \end{array}\right),~\qv_2(\theta)= 
        \left(\begin{array}{c}
          x_3(\theta) \\
          x_4(\theta) \\
      \end{array}\right)
      =\left(
      \begin{array}{c}
          \al\cos{\theta} \\
          \be\cos{\theta}
        \end{array}\right),
\end{equation}
with $(\al,\be)$ a ground state associated to the special case $\theta=0$ i.e. to the second pattern. 
We note that asymptotically for 
large $N$ we get the same ground state energy per site for each value of 
$\theta$. Because the surface on which the Hamiltonian is constant is of the 
form $\qv_1^2+\qv_2^2=C^2$, these are the only ground states.
For finite $N$, however, there are finitely many ($q(q-1)$) ground states,
corresponding to one particular value of $\theta$ (The 
exact symmetry of choosing a different pair of Potts directions gives the 
$q(q-1)$ ground states). This is an example of
chaotic size dependence, based on the breaking of a stochastic symmetry,
of the same nature as in \cite{sosy}. Because of weak compactness, different
subsequences exist whose $q(q-1)$-tuples of ground states converge to 
$q(q-1)$-tuples, associated to particular $\theta$-values. These subsequences
depend on the random pattern realization. See Appendix \ref{CSE}. 
For further background on chaotic size dependence and its role in the theory
of metastates we refer to \cite{N&S}. 
For $m\geq3$ patterns 
%and $q=3$ you have 
one has
the same discrete structure as before, but instead of a continous circle 
symmetry we have a continous $m$-sphere symmetry (isomorpic to $O(m)$). 
%If you take $q>3$ then the discrete structure of the ground states gives rise to a different labelling. This gives a group of $q(q-1)$ elements, which is some subgroup of $S_q$. 
\section{Positive temperatures
%Circleradius for nonnegative temperature 
}\label{radius} 
%{\it More things can be said about the radius of the circle through the ground states. For the Hamiltonian we use the Potts-Mattis-Hopfield Hamiltonian (\ref{HHopfield}). First we see a derivation of the fixed point equations for general $q$ and 2 pattern-variables. Then with these equations we will see we can actually calculate the radius for Gaussian patterns with Ising and $q=3$ Potts spins.}\\\\   
At positive temperatures instead of minimizing an energy one needs to 
minimize a free energy expression.\\\\
By making use of arguments from large deviation theory we obtain (see 
e.g. \cite{largedeviation}): 
\[-\be f(\be)=\sup_{\qv_1,\qv_2}\{Q(\qv_1,\qv_2)-c^{\star}(\qv_1,\qv_2)\},\]
where $f$ is the free energy per spin and $-\be H=N\frac{K}{2}(\qv_1^2+\qv_2^2)\equiv NQ$. The function $c^\star$ is the Legendre transform of $c$, where $c$ is defined as follows:
\[c(\tv)=\limNinf\frac{1}{N}\ln{\left\{\bfE_\si\exp({\vec t_1}\cdot N{\vec q_1}+{\vec t_2}\cdot N {\vec q_2})\right\}}.\]
Here ${\vec t_1}$ and ${\vec t_2}$ are vectors in $\R^{q-1}$ and $\tr_\si$ is the normalised trace at a single site. To determine the supremum (maximum)
we differentiate and put the derivative equal to $0$. This implies that
%So 
for $\qv_1$ and $\qv_2$ it holds:
\begin{equation}\label{fixed}
(\qv_1,\qv_2)_{\mbox{max}}=\nabla c(\nabla Q(\qv_1,\qv_2))=\nabla c(K \qv_1,K \qv_2)\mbox{, with}
\end{equation}
\[\left\{\begin{array}{c}
        K \qv_1=\frac{\partial Q}{\partial \qv_1} \\
        K \qv_2=\frac{\partial Q}{\partial \qv_2}.
  \end{array}\right.\]
%For this we 
We make use of the fact that for a convex function $c$, 
%holds 
$\nabla c^\star=(\nabla c)^{-1}$ (see also \cite[chap. 3]{Hopfield} and compare pag. 27). Now let us rewrite $c(\tv)$:
\[c(\tv)=\limNinf\frac{1}{N}\ln{\left\{\bfE_\si\exp({\vec t_1}\cdot N{\vec q_1}+{\vec t_2}\cdot N {\vec q_2})\right\}}=\]
\[\cdots = <\ln{\tr_\si\{\exp{(\xi {\vec t_1}+\eta {\vec t_2})\cdot\bfe^\si}\}}>_{\xi,\eta}.\]
%With
Plugging this into
(\ref{fixed}) we get the mean field equations for the order parameters 
%so called 
which have the structure of a system of
fixed point equations {\bf $\qv =F (\qv)$}:        
\begin{equation}\label{fixedpointequations}
\left\{
        \begin{array}{l}
        \qv_1=
          \left\langle
          \frac{\tr_\si\{\xi\bfe^\si\exp{[K(\xi q_1+\eta q_2)\cdot\bfe^\si]}\}} 
               {\tr_\si\{\exp{[K(\xi q_1+\eta q_2)\cdot\bfe^\si]}\}}
          \right\rangle_{\xi,\eta} \\
        \qv_2=
          \left\langle
          \frac{\tr_\si\{\eta\bfe^\si\{\exp{[K(\xi q_1+\eta q_2)\cdot\bfe^\si]}\}} 
               {\tr_\si\{\exp{[K(\xi q_1+\eta q_2)\cdot\bfe^\si]}\}} 
          \right\rangle_{\xi,\eta}.
        \end{array}
     \right.
\end{equation}
If we are in the allowed area, that is, the domain of definition of F, 
it is equivalent to look in the $(\bfe_1,\cdots,\bfe_q)$
-space.
%-plane. 
%{\bf Dit lijkt me overgedetermineerd, de orde parameter ligt in een 
%hypervlak van dimensie q-1 in de q-dimensionale ruimte
We may rewrite (\ref{fixedpointequations}) 
%there 
in this area
as follows:
\[\left\{\begin{array}{l}
        \qv_1=
                \left(
                \begin{array}{c}
                 a_1 \\
                \vdots \\
                a_q
                \end{array}
                \right)=
                        \left(
                         \begin{array}{c}
                                    \left\langle
                                 \frac{\xi\exp{K(\xi a_1+\eta b_1)}} 
                                 {\sum_{i=1}^q\exp{K(\xi a_i+\eta b_i)}}
                                  \right\rangle_{\xi,\eta} \\
                                 \vdots \\
                                 \left\langle
                                 \frac{\xi\exp{K(\xi a_q+\eta b_q)}}
                                 {\sum_{i=1}^q\exp{K(\xi a_i+\eta b_i)}}
                                \right\rangle_{\xi,\eta}
                        \end{array}
                        \right)                 \\
                                                \\
        \qv_2=
                \left(
                \begin{array}{c}
                 b_1 \\
                  \vdots \\
                  b_q
                \end{array}
                \right)=
                        \left(
                          \begin{array}{c}
                                 \left\langle
                                 \frac{\eta\exp{K(\xi a_1+\eta b_1)}} 
                                 {\sum_{i=1}^q\exp{K(\xi a_i+\eta b_i)}}
                                 \right\rangle_{\xi,\eta} \\
                                 \vdots \\
                                \left\langle
                                \frac{\eta\exp{K(\xi a_q+\eta b_q)}}
                                {\sum_{i=1}^q\exp{K(\xi a_i+\eta b_i)}}
                                \right\rangle_{\xi,\eta}
                         \end{array}
                 \right)
                 \end{array}
       \right.\]
with $\qv_1=\sum_{i=1}^q a_i\bfe_i$ and $\qv_2=\sum_{i=1}^q b_i\bfe_i$.
\subsection{Ising spins}
If we look at the behaviour for $N\ra\infty$, 
then due to the strong law of large numbers 
$\frac{1}{N}\sum_{i=1}^N \xi_i=\bfE\xi=0$. Each coordinate $a_j$ of vector 
$\qv_1=(a_1,a_2)$ is defined as 
$\frac{1}{N}\sum_{i=1}^N \xi_i\de(\si_i,\si_j)$. This means that $a_j$ is the 
contribution of the spins in the {\it j}-th direction to the sum 
$\frac{1}{N}\sum_{i=1}^N \xi_i$. Therefore: 
$a_1+a_2=\frac{1}{N}\sum_{i=1}^N \xi_i=0$ a.e. This gives a necessary 
condition for the allowed area of Ising spins:
\begin{equation}\label{con}
a_1=-a_2~\wedge~b_1=-b_2.
\end{equation}
Furthermore for 
%the 
all
Gibbs states the value of the 
%Hamiltonian 
energy
is constant, 
therefore:
\begin{equation}\label{hc}
a_1^2+a_2^2+b_1^2+b_2^2=\frac{{\rs}^2}{2}.
\end{equation}
When we substitute (\ref{con}) in equation (\ref{hc}) and project the result 
to the $(x_1,x_2)$-plane by the projection $\Pi:~\bfe_1\ra 1,~\bfe_2\ra -1$, 
we obtain the following equation:
\[x_1^2+x_2^2={\rs}^2.\]
Thus to get the radius of the circle of 
%groundstates 
Gibbs states
$\rs$, just take 
the point $\qv_1=(a,-a),~\qv_2=(0,0)$. This corresponds to the point 
$(2a,0)$ in the $(x_1,x_2)$-plane, by the projection $\Pi$. Of course 
$2a=\rs$.\\\\
With this we calculate the equation for the first coordinate of $\qv_1$ in the 
$(\bfe_1,\bfe_2)$-plane by substituting the corresponding fixed point equation:
\[a=\frac{1}{2\pi}\int\int\xi\frac{\exp{\be\xi a}}{\exp{\be\xi a}+\exp{(-\be\xi a)}}\exp{\left(-\frac{\xi^2+\eta^2}{2}\right)}d\xi d\eta=\]
\[\frac{1}{\sqrt{2\pi}}\int\xi\frac{\exp{\be\xi a}}{\exp{\be\xi a}+\exp{(-\be\xi a)}}\exp{\left(-\frac{\xi^2}{2}\right)}d\xi.\]
We replaced $K$ by $\be$, because for Ising spins $K=2\be(2-1)/2=\be$. 
The equation for the second coordinate of $\qv_1$ we calculate in the same way.
The vector $\qv_2$ is simply $(0,0)$. Now project $\qv_1$ and $\qv_2$ to the 
$(x_1,x_2)$-plane. That is done by subtracting the second coordinate of the 
$\qv_i$'s from the first one. We get the following equation for the 
radius $r^\star$:
\begin{equation}\label{rI}
r^\star=\frac{1}{\sqrt{2\pi}}\int\xi\tanh{\left(\frac{\be\xi r^\star}{2}\right)}\exp{\left(-\frac{\xi^2}{2}\right)}d\xi.
\end{equation}
For $\be>\be_0$ this equation has a nontrivial solution for $\rs$. The 
equation is the same as in \cite{sosy} except the factor 1/2 in the tanh. 
This is due to our using the Wu representation. 

\subsection{Potts spins}
If we take $q=3$, then $K=\frac{4}{3}\be$. 
The set of ground states now can be parametrized by three (in general 
${q(q-1)} \over {2}$) circles, and similarly for the low-temperature Gibbs 
states.
To 
%get here the 
obtain
the radius $\rh$ of 
%the 
such a
circle 
%of 
parametrizing the
ground or Gibbs states, we follow the same recipe as 
%with 
in the case of
Ising spins.
Here we take the 
%point
point $(\qv_1,\qv_2)$ with $\qv_1=(0,\rh/\sqrt{3},-\rh/\sqrt{3})$ and $\qv_2=(0,0,0)$ 
(the representatives of both $\qv_i$ in the \bfeplane). Now $\qv_1$ projects to $(0,\rh)$ by the projection:
\[\left(\begin{array}{c}
        x_1\\
        x_2\\
        \end{array}
  \right)=\left(
        \begin{array}{rrr}
        1 & -\frac{1}{2} & -\frac{1}{2} \\
        0 & \frac{1}{2}\sqrt{3}& -\frac{1}{2}\sqrt{3}
        \end{array}\right)
  \left(\begin{array}{c}
        a_1\\
        a_2\\
        a_3
        \end{array}\right).\]
So if we substitute the corresponding fixed point equations for $\qv_1$ {\it in the \bfeplane}, we get for 
the order parameter values
$(a_1,a_2,a_3)\equiv \qv_1$ the following 
mean field
equations:
\[\left(\begin{array}{c}
        a_1\\
        a_2\\
        a_3
        \end{array}
  \right)=\left(
        \begin{array}{c}
        0 \\
        \frac{1}{\sqrt{2\pi}}\int\xi\frac{\exp{(K\xi\rh/\sqrt{3})}}{\exp{(K\xi\rh/\sqrt{3})}+\exp{(-K\xi\rh/\sqrt{3})}+1}\exp{\left(-\frac{\xi^2}{2}\right)}d\xi \\
        \frac{1}{\sqrt{2\pi}}\int\xi\frac{\exp{(-K\xi\rh/\sqrt{3})}}{\exp{(K\xi\rh/\sqrt{3})}+\exp{(-K\xi\rh/\sqrt{3})}+1}\exp{\left(-\frac{\xi^2}{2}\right)}d\xi
        \end{array}\right).\]
Here $(a_1,a_2,a_3)=(0,\rh/\sqrt{3},-\rh/\sqrt{3})$. Thus by taking the 
difference between $a_2$ and $a_3$ and multiplying 
%that 
it
by $\frac{1}{2}\sqrt{3}$, we finally get the following expression for 
the absolute value
$\rh$:
\[\rh=\frac{1}{2\sqrt{\pi}}\sqrt{\frac{3}{2}}\int\xi\frac{\exp{(K\xi\rh/\sqrt{3})}-\exp{(-K\xi\rh/\sqrt{3})}}{\exp{(K\xi\rh/\sqrt{3})}+\exp{(-K\xi\rh/\sqrt{3})}+1}\exp{\left(-\frac{\xi^2}{2}\right)}d\xi=\]
\begin{equation}\label{rtilde}
\frac{1}{\sqrt{\pi}}\sqrt{\frac{3}{2}}\int\frac{\xi\sinh{(K\xi\rh/\sqrt{3})}}{2\cosh{(K\xi\rh/\sqrt{3})}+1}\exp{\left(-\frac{\xi^2}{2}\right)}d\xi.
\end{equation}
We can easily check that this expression indeed 
%gives 
approaches
the 
%right 
one for the
radius for the 
%circle 
circles
through the 
%groundstate
ground states, by considering the behaviour of the integrand for $K\ra\infty$. It behaves like:
\[\int |\xi|\exp{\left(-\frac{\xi^2}{2}\right)}d\xi.\]
\appendix

\section{Stochastic symmetry breaking for $q=3$}\label{CSE}
In this Appendix we adapt the fluctuation analysis of \cite{sosy} to
include Potts spins. We essentially follow the same line of argument, and 
find that the fluctuations, properly scaled, after dividing 
out the discrete symmetry, approach again a Gaussian process on the circle.\\\\
For notational simplicity we treat the case $q=3$ only. 
For $q>3$ a similar 
%analogous 
analysis applies.
%can be done 
%without new substantial difficulties.
Define the function $\phi_{N,2}$ as follows:
\[\be\phi_{N,2}(\zv)=-Q(\zv)+\zv\cdot\nabla Q(\zv)-c(\nabla Q(\zv)),\]
where $c(\tv)$ equals:
\[c(\tv)=\frac{1}{N}\ln{\left\{\bfE_\si\exp{\tv_1\cdot N\qv_1+\tv_2\cdot N\qv_2}\right\}}=\frac{1}{N}\sum_{i=1}^N\ln{\left\{\bfE_{\si_i}\exp{\tv_1\cdot\xi_i\bfe^{\si_i}+\tv_2\cdot\eta_i\bfe^{\si_i}}\right\}}.\]
This $\phi_N$ is chosen such that for $N\ra\infty$ the measure
\[{\tilde{\cal L}}=\frac{e^{-\be N \phi_N}}{Z_{N,\be}}\ra{\cal L},\]
where ${\cal L}$ is the induced distribution of the overlap parameters.\\\\
For $q=3$ it holds:
\[Q(\zv)=\frac{K}{2}\|\zv\|_2^2=\frac{2}{3}\be\|\zv\|^2.\]
Thus:
\[\phi_{N,2}(\zv)=\frac{2}{3}\|\zv\|_2^2-\frac{1}{\be N}\ln{\left\{\bfE_\si\exp \frac{4}{3}\be(\xi_i\zv_1\cdot\bfe^{\si_i}+\eta_i\zv_2\cdot\bfe^{\si_i})\right\}}\equiv\frac{2}{3}\|\zv\|_2^2-\frac{1}{\be N}\Xi_{N,2}.\]
\[\Xi_{N,2}=\sum_{i=1}^N\ln{\left\{\frac{1}{3}\exp{K(\xi_i z_{11}+\eta_i z_{21})}+\frac{2}{3}\exp{-\frac{K}{2}(\xi_i z_{11}+\eta_i z_{21})}\cosh{\frac{K\sqrt{3}}{2}(\xi_i z_{12}+\eta_i z_{22})}\right\}}=\]
\[\sum_{i=1}^N\ln{\left\{\frac{1}{3}\phi_1(z_{11},z_{22})_{\xi,\eta}+\frac{2}{3\sqrt{\phi_1(z_{11},z_{22})_{\xi,\eta}}}\phi_2(z_{12},z_{22})_{\xi,\eta}\right\}}.\]
Because for finite $N$ the set of 6 
%ground 
Gibbs
states has a discrete symmetry, as mentioned before, we choose out of these
6 states one state 
%groundstate 
%Gibbs state
we like, 
namely the one of the form $(0,\pm\al\sin{\theta},0,\pm\al\cos{\theta})$. Note that the $\theta$ depends both on $N$ and on the realization of the
random disorder variable. Then $z_{11}=z_{21}=\phi_1=0$. Inserting this and 
defining $z_{12}=\tilde{z_1}$ and $z_{22}=\tilde{z_2}$ we get for $\phi$:
\[\phi(\tilde{z_1},\tilde{z_2})=\frac{2}{3}\|(\tilde{z_1},\tilde{z_2})\|_2^2-\frac{1}{\be N}\sum_{i=1}^N\ln{\left\{\frac{1}{3}+\frac{2}{3}\cosh{\frac{2}{\sqrt{3}}\be(\xi_i\tilde{z_1}+\eta_i\tilde{z_2})}\right\}}.\]
Putting  $(z_1,z_2)=\frac{2}{\sqrt{3}}(\tilde{z_1},\tilde{z_2})$  we obtain:
\[\phi(z_1,z_2)=\frac{1}{2}\|(z_1,z_2)\|_2^2-\frac{1}{\be N}\sum_{i=1}^N\ln{\left\{\frac{1}{2}+\cosh{\be(\xi_iz_1+\eta_iz_2)}\right\}}-\frac{1}{\be N}\ln\frac{2}{3}.\]
>From now on the last term will be ignored. So it is enough to prove 
now that {\it with} the $\frac{1}{2}$ term 
we get the desired chaotic pairs structure between the patterns due to 
the quenched disorder for this class of ground states, once we  divide 
out the appropriate discrete  Potts permutation symmetry.  
%\bf Dit volg ik even niet. Bedoel je hier dat voor elke realisatie er een 
%plus-min symmetry is? In dat geval moet het er anders staan.
%And because of 
%equivalence 
%this symmetry, we get it also for all the other classes
Thus the original model displays chaotic 6-tuples.\\\\
Therefore we only need to control the fluctuations of $\phi$. Define 
\begin{equation}\label{estphi}
f_N^\star(\zv)-\bfE f_N^\star(\zv)\equiv\frac{1}{\be N}\sum_{i=1}^N\ln{\left\{\cosh{\be\zv\cdot(\xi,\eta)}\right\}}-\frac{1}{\be N}\sum_{i=1}^N\bfE\ln{\left\{\cosh{\be\zv\cdot(\xi,\eta)}\right\}}.
\end{equation}
This is the fluctuation of the Ising case which we can estimate by \cite{sosy}.
Denote the corresponding $\phi$ function by $\phi^\star$. We start with 
the following lemma:
\begin{lemma}\label{estimationphi}
\begin{equation}\label{esp}
\exp{(-\be N\phi)}\leq\exp{(-\be N\phi^\star)}.
\end{equation}
\end{lemma}
{\it Proof:}\\\\
Because
\[\exp{(-\be N\phi)}=\exp{(-\be N\bfE\phi^\star)}\exp{(-\be N(\phi-\bfE\phi^\star)},\]
we only have to estimate the quantity $\phi-\bfE\phi^\star$.
Notice that also a lower bound is essential, because the quantity can become negative. First the estimate from above:
\[\phi-\bfE\phi^\star=
\frac{1}{\be N}\sum_{i=1}^N\ln{\left\{\frac{1}{2}+\cosh{\be\zv\cdot(\xi,\eta)}\right\}}-\frac{1}{\be N}\sum_{i=1}^N\bfE\ln{\left\{\cosh{\be\zv\cdot(\xi,\eta)}\right\}}.\]
Now use
\[\ln{\left\{\frac{1}{2}+\cosh{\be\zv\cdot(\xi,\eta)}\right\}}=\ln{\left\{1+\frac{1}{2\cosh{\be\zv\cdot(\xi,\eta)}}\right\}}+\ln{\left\{\cosh{\be\zv\cdot(\xi,\eta)}\right\}}\leq\]
\[\ln{\left\{\cosh{\be\zv\cdot(\xi,\eta)}\right\}}+\ln{\frac{3}{2}}\mbox{ to get}\]
\[\phi-\bfE\phi^\star\leq\frac{1}{\be N}\sum_{i=1}^N\ln{\left\{\cosh{\be\zv\cdot(\xi,\eta)}\right\}}-\frac{1}{\be N}\sum_{i=1}^N\bfE\ln{\left\{\cosh{\be\zv\cdot(\xi,\eta)}\right\}}+\frac{1}{\be}\ln{\frac{3}{2}}=\]
\begin{equation}\label{above1}
f^\star_N(\zv)-\bfE f^\star_N(\zv)+\frac{1}{\be}\ln{\frac{3}{2}}.
\end{equation}
This is because $\cosh{x}\geq 1$ for all $x\in\R$.\\\\
The lower bound is easy because:
\[\frac{1}{\be N}\sum_{i=1}^N\ln{\left\{\frac{1}{2}+\cosh{\be\zv\cdot(\xi,\eta)}\right\}}\geq\frac{1}{\be N}\sum_{i=1}^N\ln{\left\{\cosh{\be\zv\cdot(\xi,\eta)}\right\}}.\]
This is due to the fact that the function $\ln{\al}$ is monotonically increasing in $\al$. Then it follows that:
\begin{equation}\label{under1}
\phi-\bfE\phi^\star\geq f^\star_N(\zv)-\bfE f^\star_N(\zv).
\end{equation}
Combine (\ref{above1}), (\ref{under1}) and use the fact that in the limit $\limNinf$ the constant term $\frac{1}{\be}\ln{\frac{3}{2}}$ does not 
contribute to the expression 
$\exp{\left\{-\be N (\phi-\bfE\phi^\star)\right\}}$ to 
%prove 
conclude the proof of 
lemma \ref{estimationphi}.\\\\
Henceforth it is convenient to transform  $\phi^\star$ to polar coordinates. Define $z=(r\cos{\theta},r\sin{\theta})$. Then 
$(\ref{estphi})$ transforms to:
\[|\fNb(r,\theta)|=\left|\frac{1}{\be}\bfE_\psi\bfE_\ze\ln{\cosh{\left\{\be\ze r\cos{\psi}\right\}}}-\frac{1}{\be N}\sum_{i=1}^N\ln{\cosh{\left\{\be r\ze_i\cos{(\theta-\psi_i)}\right\}}}\right|=\]
\[|\EfN^\star(r,\theta)-f^\star_N(r,\theta)|.\]
Here $\ze,\psi$ denote the polar decomposition of the two-dimensional 
vector $(\xi,\eta)$, i.e. $\ze$ is distributed with density 
$x\exp{-x^2/2}$ on $\R^+$ and $\psi$ uniformly on the circle $[0,2\pi)$. 
See \cite[page 188]{sosy}. 
This we see easily because:
\[\xi z_1+\eta z_2=(\zeta\cos{\psi})(r\cos{\theta})+(\zeta\sin{\psi})(r\sin{\theta})=\zeta r(\cos{\theta}\cos{\psi}+\sin{\theta}\sin{\psi})=\]
\[\zeta r\cos{(\theta-\psi)}\mbox{ and }\bfE_\psi\cos{(\theta-\psi)}=\bfE_\psi\cos{\psi}.\]
With $\phi^\star$ in this form, estimate (\ref{esp}) of lemma \ref{estimationphi} is not very useful, since the fluctuations of $\phi$ reach their minimum for a different radius (in $\rt$) in general than 
the fluctuations of $\phi^\star$ (in $\rs$). Thus we need to transform 
$\phi^\star$ such that the fluctuations of the 
%translated 
transformed
$\phi^\star$ 
reach their minimum at the same radius $\rt$ 
%of 
as those of
$\phi$. This we achieve as follows. There is a uniform transformation $\Pi$ which translates all the points
on the circle with radius $\rs$ centered at the origin to the circle centered at the origin with radius $\rt$, the radius of $\phi$. 
If we apply $\Pi$ to $\phi^\star(r,\theta)$ then we get $\phi^\star(r+\rs-\rt)$, the desired transformation of $\phi^\star(r,\theta)$. Now we can prove 
the next
lemma:
\begin{lemma}\label{f(r)}
For every $\ep>0$ holds: 
\begin{equation}\label{dif}
|\fNb(r,\theta)|\leq|\fNb(r+r^\star-\tilde{r},\theta)|+\ep.
\end{equation}
\end{lemma}
The constant $\rs$ is the radius of the circle parametrizing the set of
mean-field solutions
in the Ising case $(q=2)$. The constant 
$\rt$ is the radius $\rh$ in the 
Potts case $q=3$ rescaled by the factor $2/\sqrt{3}$, thus $\rt=(2/\sqrt{3})\rh$.\\\\
{\it Proof:}\\\\
We use the following estimate, which is lemma 2.5 from \cite{sosy}:
\begin{equation}\label{2.5}
|\bfE f_N^\star(r,\theta)-f_N^\star(r,\theta)|\leq\frac{\ep}{2}\mbox{ a.e. on every bounded set.}
\end{equation}
Define:
\[{\cal O}=\{\zv\in\R^2:\|\zv\|>r^\star+\de\},~{\cal O'}=\{\zv\in\R^2:\|\zv\|>{\tilde r}+\de\}.\]
Set ${\cal O}\subset{\cal O'}$ because $\rt\leq\rs$. Check this by using (\ref{rI}) and (\ref{rtilde}) and the scalingfactor $2/\sqrt{3}$ for $\rt$. Decompose ${\cal O'}$ as ${\cal O}\cup{\cal O'}\setminus {\cal O}$. Because ${\cal O'}\setminus {\cal O}$ is a finite set we can use estimate (\ref{2.5}). With the already obtained estimate for ${\cal O}$ in \cite{sosy}, (\ref{dif}) holds for all $(r,\theta)\in{\cal O'}$. Because of (\ref{2.5}) is true for all finite sets, (\ref{dif}) holds for all $(r,\theta)$.\\\\
Note that in a neighbourhood of $\tilde{r}$ it is equivalent to look in a neighbourhood of $r^\star$. Now we are able to prove the following theorem:
\begin{theorem}\label{thetheorem}
Let $\ind$ be the induced distribution of the overlap parameters and let
$m=m(\theta)=(\rt\cos{\theta},\rt\sin{\theta})$, where $\theta\in[0,\pi)$ is a uniformly distributed random variable. Then:
\[\ind_{N,\be}\stackrel{\mathcal{D}}{\ra}\frac{1}{2}\de_{m(\theta)}+\frac{1}{2}\de_{-m(\theta)}\equiv\ind_{\infty,\be}[m].\]
Furthermore, the (induced) AW-metastate is the image of the uniform distribution of $\theta$ under the measure-valued map $\theta\ra\ind_{\infty,\be}[m(\theta)]$.
\end{theorem}
First we prove
the following two lemmas:
\begin{lemma}\label{this}
For $\phi_N$ and $\xi_i$, $\eta_i$, with $i\in\N$ as defined above, there exist 
strictly positive constants $W,W',l,l'$ such that ($\rt$ is the largest solution of (\ref{rtilde}))
\[\frac{\int_{|\|\zv\|-\rt|\geq\de_N}e^{-\be N\phi_N(\zv)}d\zv}
       {\int_{|\|\zv\|-\rt|<\de_N}e^{-\be N\phi_N(\zv)}d\zv}
  \leq We^{-W N^l}\]
on a set of $\bfP$-measure at least $1-W'e^{-K'N^{l'}}$, where $\de_N=N^{-\frac{1}{10}}$. 
\end{lemma}
\begin{lemma}\label{that}
Assume the hypotheses of lemma \ref{this}. Let $a_N=N^{-1/25}$. Then there exist strictly positive constants $K_1,K_2,C_1,C_2$ such that on a set of $\bfP$-measure at least $1-K_1e^{-N^{1/25}}$ the following bound holds,
\[ \frac{\int_{A'_N}e^{-\be N\phi_N(\zv)}d\zv}
       {\int_{A_N}e^{-\be N\phi_N(\zv)}d\zv}
  \leq C_1e^{-N^{1/5}},\]
where
\[A_N=\{(r,\theta)\in\R_0^+\times [0,2\pi)||r-\rt|<\de_N,~\gnt-\gntmin<a_N\}\]
\[A'_N=\{(r,\theta)\in\R_0^+\times [0,2\pi)||r-\rt|<\de_N,~\gnt-\gntmin\geq a_N\}.\]
\end{lemma}
%With
Here
\[\gnt=\frac{\sqrt{N}}{\be}\bfE_\psi\bfE_\ze\ln{\left\{\frac{1}{2}+\cosh{\be\ze\rt\cos{\psi}}\right\}}-\frac{1}{\be\sqrt{N}}\sum_{i=1}^N\ln{\left\{\frac{1}{2}+\cosh{\left\{\be\rt\ze_i\cos{(\theta-\psi_i)}\right\}}\right\}},\]
which is the 
%polarform 
polar coordinate form
of the function $g_N(\zv)$, which is defined as:
\[g_N(\zv)=\frac{1}{\sqrt{N}}\sum_{i=1}^N\left\{\ln{\left\{\frac{1}{2}+\cosh{\be\zv\cdot(\xi,\eta)}\right\}}-\bfE\ln{\left\{\frac{1}{2}+\cosh{\be\zv\cdot(\xi,\eta)}\right\}}\right\}.\]
It is convenient to look at the following decomposition:
\[(\phi_N-\bfE\phi_N)(\zv)=\be\sqrt{N}(g_N(\zv')+h_N(\zv))\mbox{, where}\]
\[h_N(\zv)=g_N(\zv)-g_N(\zv').\]
The variable $\zv'$ is the projection of $\zv$ onto $S^1(\rt)$. Note that $\be\sqrt{N}g_N=\phi_N-\bfE\phi_N\equiv\fNbt$.
Define $g_N^\star$ and $h_N^\star$ in the same way but as decomposition of $\fNb$ instead of $\fNbt$.\\\\  
{\it Proof of lemma \ref{this}:}\\\\
Compare lemma 2.1 in \cite{sosy}. Define:
\[{\cal O}=\{\zv\in\R^2:\|\zv\|>r^\star+\de\},~{\cal O'}=\{\zv\in\R^2:\|\zv\|>{\tilde r}+\de\},\]
\[{\cal I}=\{\zv\in\R^2:\|\zv\|\leq r^\star-\de\},~{\cal I'}=\{\zv\in\R^2:\|\zv\|\leq{\tilde r}-\de\}.\]
Now we first estimate the numerator which we can also write as:
\[ \int_{|\|\zv\|-\rt|\geq\de_N}e^{-\be N\phi_N(\zv)}d\zv=\int_{{\cal O'}\cup{\cal I'}}e^{-\be N\bfE\phi_N(\zv)}e^{-\be N(\phi_N(\zv)-\bfE\phi_N(\zv))}.\]
With lemma \ref{f(r)} we have the following inequality:
\[\sup_{\zv\in{\cal O'}}|\fNb(r,\theta)|-\ep\leq\sup_{\zv\in{\cal O'}}|\fNb(r+r^\star-\tilde{r},\theta)|=\sup_{\zv\in{\cal O}}|\fNb(r)|\]
\begin{equation}\label{theCtrick}
\bfP\left[\sup_{(r,\theta)\in{\cal O'}}|\fNb(r,\theta)|-\ep\geq\frac{C}{2}(r-\rt)^2\right]\leq\bfP\left[\sup_{(r,\theta)\in{\cal O}}|\fNb(r,\theta)|\geq\frac{C}{2}(r-r^\star)^2\right].
\end{equation}
Lemma 2.4 of \cite{sosy} tells us that this event is of measure zero. Now we can estimate the integral.\\\\
First we estimate $\bfE\phi^\star_N(\zv)$. Because $\bfE\phi^\star_N(\zv)$ is a bounded 
function, in each bounded interval one can always bound 
%$\bfE\phi_N(z)$ 
it
from below by a function of 
%this 
the following
kind:
\[\bfE\phi^\star_N(\zv)\geq C(\|\zv\|-\rt)^2+\bfE\phi^\star_N(\rt),\mbox{ with $C$ a positive bounded constant.}\]
Then 
\[\bfE\phi^\star_N(\zv+\rs-\rt)\geq C(\|\zv\|-\rs)^2+\bfE\phi_N(\rs),\]
when we apply $\Pi$ to this estimate. Now use 
%estimation 
estimate
(\ref{theCtrick}) with {\it this} constant $C$. Then 
it
holds:
\[\int_{\cal O'}e^{-\be N\bfE\phi_N(\zv)}e^{-\be N(\phi_N(\zv)-\bfE\phi_N(\zv))}dz\leq\int_{\cal O}e^{-\be N\bfE\phi^\star_N(\zv+\rs-\rt)}e^{\be N|\fNb(\zv)|}e^{\ep\be N}d\zv\leq\]
\[e^{-\be N(\bfE\phi^\star_N(\rs)-\ep)}\int_{\cal O}e^{-\be NC(r-\rs)^2}e^{\be N \frac{C}{2}(r-\rs)^2}dr=e^{\be N(\bfE\phi^\star_N(\rs)-\ep)}\int_{\cal O}e^{-\be N\frac{C}{2}(r-\rs)^2}dr\leq\]
\[\cdots\leq 2\pi\frac{2}{\be NC}\exp{\left(-\be N(\bfE\phi^\star_N(\rs)-\ep)\right)}\exp{-\be N C_2\left(\frac{\de^2}{4}\right)}.\]
For further details see \cite{sosy}. The interior ${\cal I}$ gives a similar expression. Notice that the image of ${\cal I}$ under the transformation $\Pi$: $r\ra r+\rs-\rt$ is ${\cal I}\setminus B(0,\rs-\rt)$. The ball $B(0,\rs-\rt)$ is a finite set so we can integrate over ${\cal I}$ instead of ${\cal I}\setminus B(0,\rs-\rt)$ by (\ref{2.5}). 
To estimate the denominator we just replace $\rs$ by $\rt$ in the expressions of the proof of lemma 2.1 in \cite[pag.192,193]{sosy}. Combining the estimates for the numerator and the denominator gives the desired result.\\\\
{\it Proof of lemma \ref{that}:}\\\\
>From this moment we ignore the constants which enter by applying lemma \ref{estimationphi}, because they cancel out when we divide the numerator by the denominator. For $|h_N|$ it holds:
\[|h_N|\leq|h^\star_N|\leq\ep,\]
by lemma 2.6 of \cite{sosy}. Consider the following integral:
\[\int_{\theta:\gnt>a_N+\gntmin}e^{-\sqrt{N}\gnt}\leq 2\pi e^{-\sqrt{N}\gntmin}e^{\sqrt{N}a_N}.\]
Henceforth it is just a matter of plugging in to get the desired estimate on the denominator. We refer to \cite{sosy} for the details. One gets a 
estimate for the denominator in the same way. By dividing the two estimates
lemma \ref{that} is proven.\\\\
{\it Proof of theorem \ref{thetheorem}:}\\\\
In the preceding paragraphs we have seen that the measures ${\tilde{\cal L}}$ concentrate on a circle at the places where the random function $\gnt$ takes its minimum. Now it only remains to show that these sets degenerate to a single point, a.s. in the limit $N\ra\infty$. If we have proven it for ${\tilde{\cal L}}$, then we have proven it also for ${\cal L}$, because $\limNinf {\tilde{\cal L}}={\cal L}$. With the help of $\cite{sosy}$ this is very easy, because we can use Proposition 3.4 with the function
\[g(.)=\ln{\left\{\cosh{\be.}+\frac{1}{2}\right\}}.\]
This works because $g$ is an aperiodic even function.  
And of course Proposition 3.7 also holds for this $g$. These two propositions we use, tell us that the process $\eta_N=\gnt-\bfE\gnt$ converges to a strictly stationary Gaussian process, having a.s. continuously differentiable sample paths. And on any interval $[s,s+t],~t<\pi$ the function $\eta_N$ has only one global minimum. Furthermore, if we define the sets:
\[L_N=\{\theta\in[0,\pi):~\eta_N(\theta)-\mbox{min}_{\theta '}\eta_N(\theta ')\leq\ep_N\},\]
with $\ep_N$ some sequence converging to zero, $L_N\stackrel{\cal D}{\ra}\theta^{\star}$.
Then the remarks below {\it Proof of theorem 3} in \cite{sosy} conclude the proof.\\\\ 
This research was supported by the Samenwerkingsverband Mathematische Fysica. A.v.E. thanks Christian Borgs for asking 
the question how the results of \cite{sosy} generalize to Potts spins.


\begin{thebibliography}{mmmm}
\bibitem[vEHP]{colour} A.C.D. van Enter, J.L. van Hemmen, and C. Pospiech, {\em Mean-field theory of random-site q-state Potts models}, 
%SFB 123 preprint nr. 433 (1987)  
J. Phys. A {\bf 21}, 791--801 (1988).
\bibitem[BvEN]{sosy} A. Bovier A.C.D. van Enter, B. Niederhauser, {\em Stochastic symmetry-breaking in a Gaussian Hopfield model}, J. Stat. Phys. {\bf 95}, 
%Nos. 1/2 
181--213
(1999).
\bibitem[B]{lecturenotes} A. Bovier, {\em Statistical mechanics of disordered systems}, MaPhySto Lecture Notes 10, Aarhus (2001).
\bibitem[BG]{Hopfield} A. Bovier, V. Gayrard, {\em Hopfield models as generalized random mean field models}, in Mathematical Aspects of spin glasses and neural networks, Progress in Probability 41, Birkh\"auser, Boston, (1997).
\bibitem[G]{G} V. Gayrard, {\em Thermodynamic limit of the q-state 
Potts-Hopfield model with infinitely many patterns}, J. Stat. Phys. 68, 
977-1011 (1992).
\bibitem[HvEC]{largedeviation} J.L. van Hemmen, A.C.D. van Enter, and J. Canisius, {\em On a classical spin glass model}, Z. Phys. B - Condensed matter 50, 311-336 (1983).
\bibitem[MPV]{MPV} M. M\'ezard, G. Parisi and M.`A.~Virasoro,
{\em Spin Glass Theory and Beyond}, World scientific, (1987).
\bibitem[Nie]{nie} B. Niederhauser, {\em Mathematical Aspects of Hopfield 
Models},  Dissertation TU-Berlin, 2000.
\bibitem[NS]{N&S} C.M. Newman and D.L. Stein, {\em Thermodynamic Chaos and the 
structure of short-range spin glasses}, in Mathematical Aspects of spin 
glasses and neural networks, Progress in Probability 41, Birkh\"auser, Boston, (1997).
\bibitem[NS2]{NS2} C.M. Newman and D.L. Stein, {\em The State(s) of 
Replica Symmetry Breaking: Mean Field Theories vs Short-Ranged Spin Glasses}, 
formerly known as {\em Replica Symmetry Breaking's New Clothes}, J. Stat.Phys.
to appear, cond-mat{/}0105282 (2001). 
\bibitem[Wu]{Potts} F.Y. Wu, {\em The Potts-model}, Rev. Mod. Phys. {\bf 54}, 235 (1982).


\end{thebibliography}
\end{document}